\pdfoutput=1

\documentclass{emulateapj}

\usepackage{amsmath}

\newcommand{\Msun}{\ensuremath{\mathrm{M}_{\odot}}}
\newcommand{\Rsun}{\ensuremath{\mathrm{R}_{\odot}}}

\newcommand{\Mstar}{\ensuremath{M_{\ast}}}
\newcommand{\Rstar}{\ensuremath{R_{\ast}}}

\newcommand{\mstar}{\ensuremath{m}}
\newcommand{\dmstar}{\ensuremath{\delta m}}

\newcommand{\Mmag}{\ensuremath{M_{\rm mag}}}
\newcommand{\Masymp}{\ensuremath{M_{\rm \infty}}}

\newcommand{\rhomax}{\ensuremath{\rho_{\rm max}}}

\newcommand{\Teff}{\ensuremath{T_{\rm eff}}}
\newcommand{\logg}{\ensuremath{\log g}}

\newcommand{\Ocrit}{\ensuremath{\Omega_{\rm c}}}

\newcommand{\soriab}{$\sigma$~Ori~AB}
\newcommand{\sorid}{$\sigma$~Ori~D}
\newcommand{\sorie}{$\sigma$~Ori~E}

\newcommand{\MOST}{\textit{MOST}}

\begin{document}

\title{\MOST\ Observations of $\sigma$~Ori~E: Challenging the Centrifugal Breakout Narrative}
\shorttitle{\MOST\ Observations of $\sigma$~Ori~E: Challenging the Centrifugal Breakout Narrative}


\author{R. H. D. Townsend\altaffilmark{1}, 
        Th. Rivinius\altaffilmark{2},
        J. F. Rowe\altaffilmark{3},
        A. F. J. Moffat\altaffilmark{4},
        J. M. Matthews\altaffilmark{5},
        D. Bohlender\altaffilmark{6},
        C. Neiner\altaffilmark{7},
        J. H. Telting\altaffilmark{8},
        D. B. Guenther\altaffilmark{9},
        T. Kallinger\altaffilmark{10,5},
        R. Kuschnig\altaffilmark{10,5},
        S. M. Rucinski\altaffilmark{11},
        D. Sasselov\altaffilmark{12},
        W. W. Weiss\altaffilmark{10}}

\email{townsend@astro.wisc.edu}

\altaffiltext{1}{Department of Astronomy, University of Wisconsin-Madison, 2535 Sterling Hall, 475 N. Charter Street, Madison, WI 53706, USA}
\altaffiltext{2}{ESO - European Organisation for Astronomical Research in the Southern Hemisphere, Casilla 19001, Santiago 19, Chile}
\altaffiltext{3}{NASA Ames Research Center, Moffett Field, CA 94035, USA}
\altaffiltext{4}{D\'epartment de physique, Universit\'e de Montr\'eal C.P. 6128, Succursale Centre-Ville, Montr\'eal, QC H3C 3J7, Canada}
\altaffiltext{5}{Department of Physics and Astronomy, University of British Columbia, 6224 Agricultural Road, Vancouver, BC V6T 1Z1, Canada}
\altaffiltext{6}{National Research Council of Canada, Herzberg Institue of Astrophysics, 5071 West Saanich Road, Victoria, BC, V9E 2E7, Canada}
\altaffiltext{7}{LESIA, UMR 8109 du CNRS, Observatoire de Paris, UPMC, Universit\'e Paris Diderot, 5 place Jules Janssen, 92195, Meudon Cedex, France}
\altaffiltext{8}{Nordic Optical Telescope, Apartado 474, 38700 Santa Cruze de La Palma, Spain}
\altaffiltext{9}{Department of Astronomy and Physics, St. Mary’s University, Halifax, NS B3H 3C3, Canada}
\altaffiltext{10}{University of Vienna, Institute for Astronomy, T\"urkenschanzstrasse 17, A-1180 Vienna, Austria}
\altaffiltext{11}{Dept. of Astronomy and Astrophysics, University of Toronto, 50 St George Street, Toronto, ON M5S 3H4, Canada}
\altaffiltext{12}{Harvard-Smithsonian Center for Astrophysics, 60 Garden Street, Cambridge, MA 02138, USA}

\shortauthors{Townsend et al.}


\begin{abstract}
We present results from three weeks' photometric monitoring of the
magnetic helium-strong star \sorie\ using the
\MOST\ microsatellite. The star's light curve is dominated by
twice-per-rotation eclipse-like dimmings arising when magnetospheric
clouds transit across and occult the stellar disk. However, no
evidence is found for any abrupt centrifugal breakout of plasma from
the magnetosphere, either in the residual flux or in the depths of the
light minima. Motivated by this finding we compare the observationally
inferred magnetospheric mass against that predicted by a breakout
analysis. The large discrepancy between the values leads us to argue
that centrifugal breakout does not play a significant role in
establishing the magnetospheric mass budget of \sorie.
\end{abstract}



\section{Introduction} \label{sec:intro}

The B2Vpe star \sorie\ (HD~37479) is a magnetic helium-strong star
characterized by variations in many of its observables, including
photometric indices \citep{Hes1977}, H$\alpha$ emission
\citep{Wal1974,Bol1974,Rei2000}, photospheric and wind absorption
lines \citep{PedTho1977,GroHun1982,ShoBro1990}, radio emission
\citep{LeoUma1993}, linear continuum polarization
\citep{KemHer1977,Car2013} and circular line polarization
\citep{LanBor1978,Oks2012}.  The variability originates from surface
abundance inhomogeneities, together with plasma trapped in a
circumstellar magnetosphere with the highest densities in co-rotating
cloud-like structures situated at the intersections between magnetic
and rotational equators
\citep[e.g.][]{GroHun1982,Bol1987,Sho1993,Tow2005}. \citet{Tow2010}
recently discovered that the $1.19\,{\rm d}$ rotation period is
gradually lengthening due to magnetic braking.

Building on previous work by \citet{Nak1985}, \citet{TowOwo2005}
developed a \emph{rigidly-rotating magnetosphere} (RRM) model to
explain the shape of the star's magnetosphere. Radiatively driven wind
streams flowing up from the photosphere are channeled into head-on
collisions by closed magnetic loops.  After shock heating and
subsequent radiative cooling the near-stationary plasma settles into
magnetohydrostatic equilibrium, supported against the inward pull of
gravity by the centrifugal force arising from enforced
co-rotation. The predicted plasma distribution appears to be in good
agreement with observations \citep[][hereafter T05]{Tow2005}, although
there are some discrepancies \citep[e.g.,][]{Car2013} which warrant
further investigation.

For such a wind-fed magnetosphere, the total mass of trapped plasma
necessarily must grow with time unless a countervailing mass leakage
mechanism allows some kind of balance to be
reached. \citet{TowOwo2005} proposed a mechanism involving the
stressing and eventual breaking of magnetic loops by the centrifugal
force, which grows in strength as plasma
accumulates. Magnetohydrodynamic (MHD) simulations by \citet{udD2006}
support this \emph{centrifugal breakout} hypothesis, and moreover
suggest that the reconnection heating arising during breakout episodes
could explain the X-ray flares seen in \sorie\ (over and above its
quiescent wind-shock emission) by \citet{GroSch2004} and
\citet{San2004}. However, no direct evidence of breakout has so far
been found.

In this paper we present data from three weeks' photometric monitoring
of \sorie\ by the \MOST\ microsatellite \citep{Wal2003}, beginning
November 2007. The motivation for this observing campaign was to
better characterize the star's light curve, and to search for any
cycle-to-cycle changes arising from putative centrifugal breakout
episodes.  Section~\ref{sec:obs-reduce} describes the observations and
explains the procedure used to reduce the raw data, and
Section~\ref{sec:analysis} analyzes various aspects of the light
curve. The findings are discussed in Section~\ref{sec:discuss} and then
summarized in Section~\ref{sec:summary}.


\section{Observations and Data Reduction} \label{sec:obs-reduce}

\MOST\ observed \sorie\ and four other nearby bright B-type stars
(HD~37525; \sorid; HD~37744; HD~294272) over the interval November 12
-- December 3 2007 with a cadence of around 60\,s. The satellite
operated in direct imaging mode, where targets are placed on the open
area of the science CCD not covered by the Fabry microlens array
\citep[see][]{Row2006a,Row2006b}; this comes at the cost of a degraded
instrumental stability and precision, but is necessary because
\sorie\ is too faint ($V=6.66$) to observe in Fabry mode. Individual
subexposures of 0.530\,s were co-added onboard the satellite prior to
downloading, to avoid saturating the telemetry link
\citep[see][]{Row2008}. The number of subexposures per co-added
exposure was initially set at 31, but was then increased to 61 after
the first 17 hours of the run.

\begin{figure*}[ht]
\includegraphics{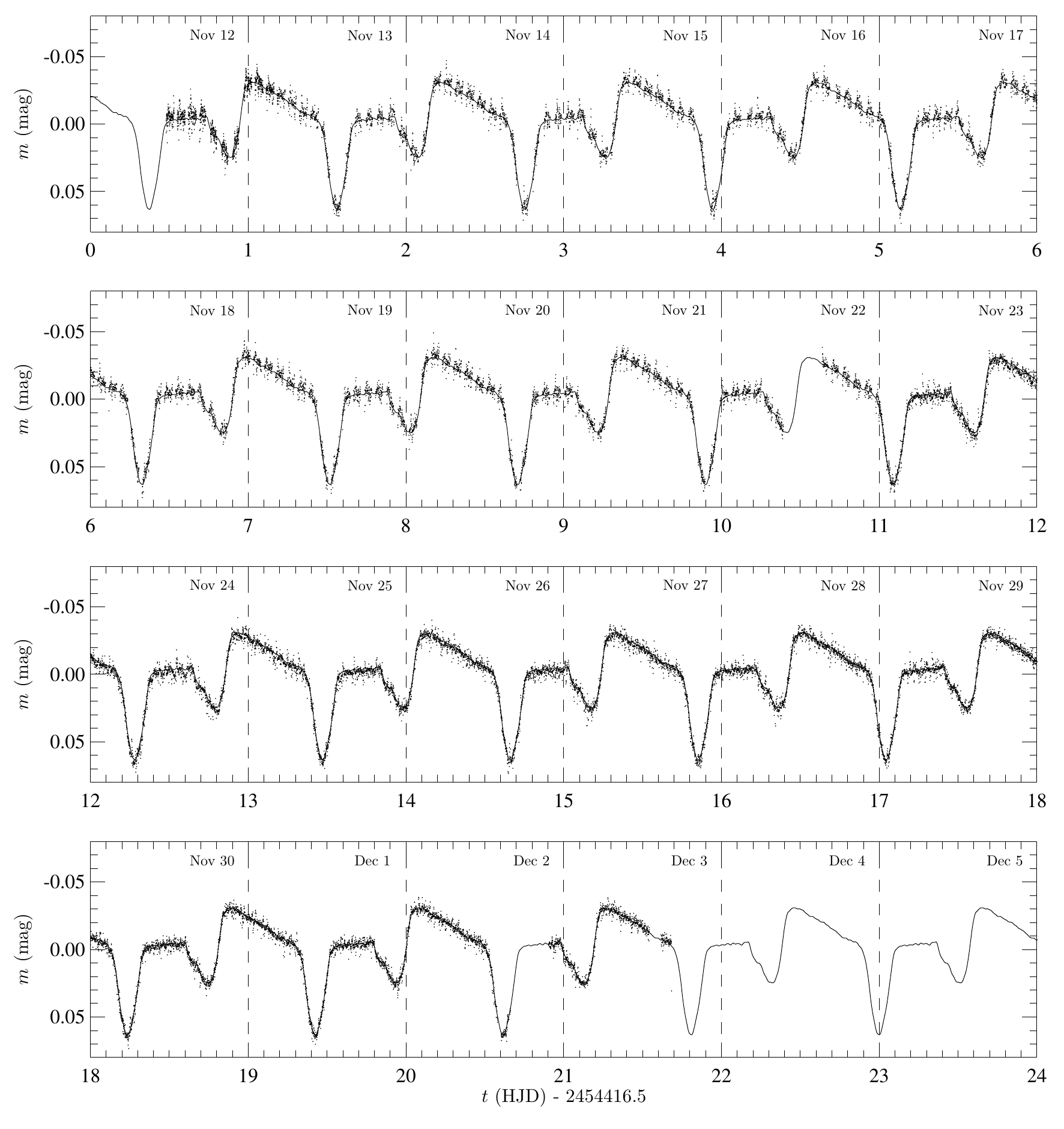}
\caption{The stellar flux \mstar\ of \sorie, in magnitudes relative to
  the mean flux, plotted as a function of time. The solid curve
  overlaying the data shows the periodic signal used for
  pre-whitening. The vertical dashed lines delineate the day
  boundaries.} \label{fig:light-curve}
\end{figure*}

At the beginning of the run \sorie\ fell outside the \MOST\ continuous
viewing zone (CVZ); therefore, for $\sim 25$ minutes of every
101.413-minute orbit the satellite slewed to observe an alternative
field in the Hyades, resulting in periodic gaps in the data (see the
top two rows of Fig.~\ref{fig:light-curve}). On November 23 the star
entered the CVZ and \MOST\ switched to observing it
continuously. Around half a day prior to this switch the onboard
computer crashed, leading to a $\sim 0.25\,{\rm d}$ gap in the
data. The orientation of the spacecraft after the switch initially led
to increased solar heating and a climb in the CCD temperature,
accounting for certain features in the residual light curve discussed
below. Finally, gaps in the data on December 2 and December 3 arose
due to science data buffer overruns.

The co-added exposures of \sorie, each a 20 by 20 pixel image, are
reduced using the standard approach of synthetic aperture
photometry. The stellar flux is calculated as the difference between
the total flux in a 5-pixel radius circular aperture centered on the
2-dimensional Gaussian centroid of the image, and the estimated
background flux. A complication peculiar to \MOST's direct imaging
mode is that the background flux includes stray light contributions
which are spatially inhomogeneous and modulate with the satellite's
orbit \citep{Ree2006}. To remove these artifacts we follow the
procedure described by \citet{Row2006a,Row2006b} with some
modifications. The correlation between the pre-whitened stellar flux
and the background flux is fit using locally weighted regression
\citep{Cle1979}, with a tri-cubed weight function and a smoothing
parameter $f=0.084$ chosen by 10-fold cross validation
\citep{ArlCel2010}. The pre-whitening subtracts a periodic signal
representing the intrinsic variability of \sorie, which would
otherwise distort the flux correlation fit. To determine this signal
we apply locally weighted regression to the phase-folded stellar flux,
with a smoothing parameter $f=0.019$ again determined by cross
validation and a period $P = 1.190847\,{\rm d}$ chosen by minimizing
the weighted mean square error of the regression. The 68.2\%
confidence interval of this period determination is $\Delta P =
\pm0.000015\,{\rm d}$ \citep[determined via bootstrap Monte-Carlo
  simulations; e.g.,][]{Pre1992}, and so the period is in good
agreement with the $P = 1.198051 \pm 0.000003 \,{\rm d}$ predicted by
the \citet{Tow2010} ephemeris.

Fig.~\ref{fig:light-curve} plots the light curve resulting from this
reduction process, together with the periodic signal determined for
the pre-whitening. These data clearly reveal the signature
twice-per-rotation eclipse-like dimmings of the star arising when the
magnetospheric clouds transit across and occult the stellar
disk. Allowing for the different photometric responses, no gross
differences stand out between the \MOST\ light curve and historical
observations \citep[e.g.,][]{Hes1977,PedTho1977,GroHun1982}.


\section{Analysis} \label{sec:analysis}

\begin{figure*}[ht]
\includegraphics{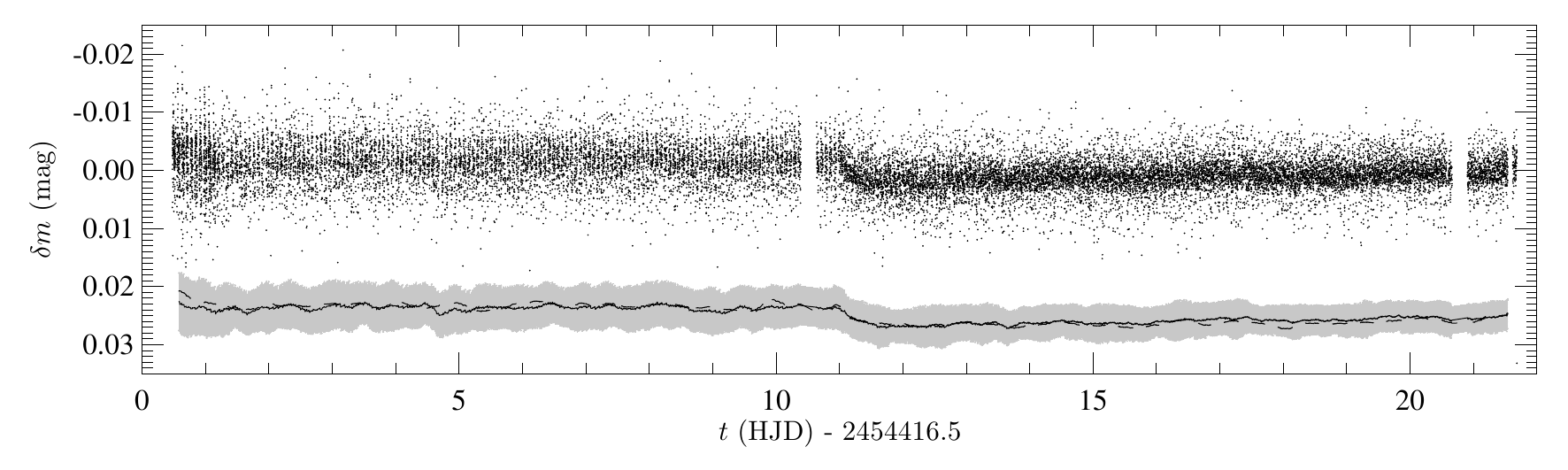}
\caption{The residual stellar flux \dmstar, in magnitudes relative to
  the mean flux, plotted as a function of time. The solid curve
  beneath (shifted down by $0.025\,{\rm mag}$ for clarity) shows the
  $\pm0.1\,{\rm d}$ boxcar mean curve, with the gray envelope
  illustrating the associated one-standard-deviation bounds. For
  comparison, the dashed curve shows the corresponding smoothed light
  curve of HD~37744.} \label{fig:residuals}
\end{figure*}

Figure~\ref{fig:residuals} shows the residual flux after pre-whitening
the light curve with the periodic signal. A $\pm0.1\,{\rm d}$ boxcar
mean curve, together with the associated one-standard-deviation
bounds, is plotted below the points to highlight long-term trends in
the data. This smoothed curve clearly reveals an abrupt dimming by
about $0.0035\,{\rm mag}$ near the mid-point of the observations ($t -
2454416.5 \approx 11\,{\rm d}$), together with a reduction in the standard
deviation. Also visible in the curve is a low-level ripple with a
frequency $\sim 1-2\,{\rm d^{-1}}$.  However, the corresponding
smoothed light curve of HD~37744 (also shown in the figure) reveals
similar behavior in both respects; hence, neither the dimming nor the
ripple can be intrinsic to \sorie. The effects are likely instrumental
in origin; the dimming in particular is correlated with a sharp
$5\,{\rm K}$ increase in the temperature of the CCD pre-amplifier, due
to the increased solar heating which occurred when \MOST\ switched to
continuous observation of \sorie\ (see Sec.~\ref{sec:obs-reduce}).

Apart from these instrumental variations, the smoothed curve in
Fig.~\ref{fig:residuals} is relatively devoid of features. In
particular, there are no obvious flares characterized by a sudden
brightening of the star followed by a slow decline. One interpretation
of this result is that there were no centrifugal breakout episodes
during the \MOST\ run, since any breakout would be accompanied by a
large release of magnetic energy. A caveat, however, is that although
a link between magnetic reconnection and optical flaring has been
established in other types of systems (e.g., weak-line T Tauri stars
--- \citealp{Fer2004}; M dwarfs --- \citealp{Ste2006}), the same
cannot be said for the centrifugally supported magnetospheres
considered here. The MHD breakout simulations by \citet{udD2006}
cannot offer much guidance, since they are unable to predict how much
emission will be produced at optical wavelengths.

In addition to flaring, centrifugal breakout episodes might reveal
themselves through abrupt and ongoing reductions in the magnetospheric
column density. To search for these signatures we measure the depths
of the primary and secondary minima in the light curve, across the 20
rotation cycles spanned by the observing run. While the depths show
cycle-to-cycle changes at a level $\sim 0.002\,{\rm mag}$ which
exceeds the formal error bars, these variations occur in both
directions and appear more consistent with the instrumental variations
mentioned in the previous section than with any evolution in the
column density.


\section{Discussion} \label{sec:discuss}

\begin{table}
\begin{tabular}{cccccc}
\Mstar (\Msun) & \Rstar (\Rsun) & $\Omega$ ( \Ocrit) & $B_{\ast}$ (kG) & $\beta$ ($\degr$) & $\epsilon_{\ast}$ \\ \hline
8.30 & 3.77 & 0.454 & 11.0 & 55 & $10^{-3}$
\end{tabular}
\caption{Parameters adopted in calculating the inferred magnetosphere
  mass \Mmag\ and asymptotic magnetosphere mass \Masymp\ of
  \sorie.} \label{tab:params}
\end{table}

The failure to find any evidence for centrifugal breakout episodes,
either in the form of optical flares in the residual flux or as
systematic changes in the depths of the light minima, could be due
simply to unlucky scheduling of the \MOST\ run coupled with the fact
that the breakout recurrence timescale is poorly constrained (as it
depends on the unknown wind mass-loss rate). However, there are a
number of independent arguments which favor the alternative conclusion
that centrifugal breakout simply \emph{does not occur} in \sorie, at
least at a level where it has any impact on the magnetospheric mass
budget.

Foremost amongst these is the discrepancy between the magnetospheric
mass \Mmag\ inferred from analysis of the observations using the RRM
model, and the asymptotic magnetosphere mass \Masymp\ predicted by the
breakout analysis of \citet[][their Appendix A2]{TowOwo2005}; if
centrifugal breakout plays a role in governing the magnetospheric mass
budget then these two values should be
comparable. Table~\ref{tab:params} lists the stellar and magnetosphere
parameters adopted here to evaluate \Mmag\ and \Masymp; the field
strength $B_{\ast}$, magnetic obliquity $\beta$ and magnetosphere
scale-height parameter $\epsilon_{\ast}$ are taken from T05, while the
other parameters are derived in Appendix~\ref{app:params}. Applying
the light-curve synthesis procedure described by T05, the RRM model
requires $\rhomax \kappa \Rstar \approx 7$ to reproduce the observed
depth $\approx 0.065\,{\rm mag}$ of the primarily light minima,
with \rhomax\ being the maximum mass density in the magnetosphere,
$\kappa$ the flux-mean opacity in the \MOST\ passband, and \Rstar\ the
stellar radius. A lower limit on the opacity is given by the electron
scattering value, $\kappa_{\rm es} = 0.34\,{\rm cm^{2}\,g^{-1}}$ for a
fully ionized solar-abundance composition. With $\Rstar = 3.77\,\Rsun$ from
Table~\ref{tab:params}, we therefore obtain an upper limit $\rhomax
\lesssim 8 \times 10^{-11}\,{\rm g\,cm^{-3}}$ on the maximum
density. Integrating over the RRM density distribution leads to a
corresponding upper mass limit $\Mmag \lesssim 2 \times
10^{-10}\,\Msun$. This is almost two orders of magnitude smaller than
the asymptotic mass $\Masymp = 1.2 \times 10^{-8}\,\Msun$ predicted by
the breakout analysis, indicating that the magnetosphere is well short
of the level required for \emph{significant} breakout episodes to
occur.

With hindsight, this result didn't have to wait for the
\MOST\ observations presented here. Certainly, these observations
provide an unprecedentedly precise characterization of the (remarkably
unchanging) light curve of \sorie, which provokes our re-examination
of centrifugal breakout. However, the same general conclusions will be
reached if a similar analysis is applied to the original
\citet{Hes1977} light curve (or for that matter any other photometric
observations of the star), since the depths of the minima in these
historic data are similar to those in Fig.~\ref{fig:light-curve}.  We
also note that the spectroscopic measurements by \citet{GroHun1982}
independently indicate $\Mmag \approx 10^{-10}\,\Msun$, and the recent
linear polarization measurements by \citet{Car2013} likewise give
$\Mmag \approx 2 \times 10^{-11}\,\Msun$ --- both consistent with the
upper limit derived above. Presumably the larger figure derived by
\citet{GroHun1982} results from their assumption of a vertical
magnetosphere extent $\sim 1\,\Rstar$, rather larger than the $\sim
0.2\,\Rstar$ predicted by the RRM model.

Further corroborating arguments against breakout are presented in a
forthcoming paper (Townsend et al., in preparation), which
demonstrates that the low-mass companion discovered by \citet{Bou2009}
is responsible for the majority of the X-ray flux from the
\sorie\ system. It seems likely that the X-ray flares proposed to
arise during breakout (Sec.~\ref{sec:intro}) instead come from the
magnetic activity of the companion, as originally conjectured by
\citet{San2004}.

These findings challenge a prevailing narrative for mass leakage from
centrifugally supported magnetospheres. It is natural to now ask what
other leakage mechanism(s) might be at work to balance the continual
feeding of plasma from the wind, as evidenced by the star's
rotationally modulated UV absorption lines
\citep{ShoBro1990}. \citet{HavGoe1984} explore cross-field diffusive
processes such as ambipolar diffusion, but find them far too slow to
be effective; revisiting their calculations with updated stellar
parameters does not change this conclusion. A related question
concerns the process(es) responsible for magnetospheric features not
predicted by the RRM model --- for instance, the substructure seen in
the secondary light minima in Fig.~\ref{fig:light-curve}, and the
departures from the expected mass distribution revealed in the linear
polarization measurements by \citet{Car2013}. Corresponding departures
can also be seen in photometric and spectroscopic observations of a
number of other He-strong stars harboring magnetospheres (e.g.,
HR~7355 --- \citealp{Oks2010,Riv2010,Riv2012}; $\delta$~Ori~C ---
\citealp{Leo2010}; HR~5907 --- \citealp{Gru2012}).  Are these a
consequence of the as-yet-unidentified mass-leakage mechanism, or
instead due to a non-dipole field topology? In the case of \sorie,
recent spectropolarimetric measurements by \citet{Oks2012} indeed
reveal deviations from dipolarity, although these are not consistent
with the decentered dipole invoked by T05 to explain the overall
difference in the depths of the primary and secondary light
minima. Clearly, there remains much work to be done in understanding
the effects of mass redistribution, mass leakage and field topology in
governing the distribution and overall amount of plasma in these
stars' magnetospheres.

Looking toward the future, a logical next step is to decompose the
\MOST\ light curve into magnetospheric and photospheric components,
the latter arising from the inhomogeneous abundance distribution
across the stellar surface. \citet{Krt2007,Krt2011} have successfully
used surface abundance maps derived from Doppler imaging to reproduce
the photospheric light variations of other He-strong stars. A similar
approach should be possible for \sorie, once the process of deriving
the abundance maps is complete \citep[see][]{Oks2012}. The decomposed
light curve will allow quantitative testing of the hypothesis (e.g.,
T05) that the brightening seen after the secondary minima is
photospheric rather than magnetospheric in origin. Likewise, comparing
the magnetospheric component against the light-curve morphologies
predicted by the RRM model \citep[see][]{Tow2008} will allow further
refinement of the model and moreover offer insights into the
as-yet-unknown mechanisms responsible for mass leakage.


\section{Summary} \label{sec:summary}

We have presented new photometric observations of \sorie\ obtained
using the \MOST\ microsatellite (Sec.~\ref{sec:obs-reduce}). Despite
the unprecedented precision of the light curve no evidence is found
for centrifugal breakout episodes or any other variability beyond
rotational modulation, either in the residual flux or in the depths of
the light minima (Sec.~\ref{sec:analysis}). Motivated by this finding
we compare the observationally inferred magnetospheric mass against
the asymptotic mass predicted by the \citet{TowOwo2005} breakout
analysis (Sec.~\ref{sec:discuss}). The former is around two orders of
magnitude smaller than the latter, leading us to rule out centrifugal
breakout as a mechanism for significant magnetospheric mass leakage in
\sorie.


\acknowledgements RHDT acknowledges support from NSF awards
AST-0908688 and AST-0904607, and NASA award NNX12AC72G. AFJM, DBG, JMM
and SMR are grateful for financial support from NSERC (Canada). RK and
WW acknowledge support by the Austrian Science Fund, P22691-N16.

\appendix

\section{Fundamental Parameters of \sorie} \label{app:params}

\citet{GroHun1982} determine an effective temperature $\Teff =
22\,500\,{\rm K}$ for \sorie\ by fitting the spectral energy
distribution from UV through to IR. They likewise derive a surface
gravity $\logg = 3.85\,{\rm dex}$ from modeling H and He equivalent
widths. A subsequent more-detailed analysis of Balmer-line wings led
\citet{Hun1989} to revise this value slightly upwards, to $\logg =
3.95\,{\rm dex}$. As discussed by these latter authors, the \Teff\ and
\logg\ together imply that \sorie\ is more distant ($\sim 650\,{\rm
  pc}$) than the $\sigma$~Ori cluster ($\sim 450\,{\rm pc}$), and is
moreover a factor $\sim 10$ older than the cluster. These findings,
however, stand contrary to a number of observational results
indicating that \sorie\ is a bona fide member of the cluster rather
than a background star. The reddening of \sorie\ is the same as
\soriab\ \citep{She2008}, and likewise for the interstellar
polarization \citep{KemHer1977,Car2013}. The radial velocity and
proper motion of \sorie\ are indistinguishable from those of the
cluster \citep{Cab2007}. Finally, the spindown measurements by
\citet{Tow2010} indicate that the star is young, with an age $\sim
1.1\,{\rm Myr}$ consistent with lower-end age estimates for the
cluster.

The problem with the \citet{Hun1989} analysis likely resides in the
surface gravity determination. Emission from magnetospheric plasma
fills in the wings of Balmer lines; if not properly corrected this
makes the lines appear less broad, and the gravity consequently
smaller, than is actually the case. Given this complication it seems
better to avoid the gravity measurement altogether, and derive stellar
parameters using a different approach. Accordingly, assuming
\sorie\ is a cluster member, a radius $\Rstar = 3.77\,\Rsun$ follows from
the angular diameter $\theta = 0.079\,{\rm mas}$ \citep{GroHun1982}
and the cluster distance $d = 444\,{\rm pc}$ derived for solar
metallicity by \citet{She2008}.

To obtain the corresponding mass, we calculate a sequence of
solar-metallicity evolutionary tracks with masses $\Mstar = 7, 7.1, 7.2,
\ldots, 9.9, 10\,\Msun$ using the MESA stellar evolution code
\citep{Pax2011}. For simplicity the calculations neglect the effects
of rotation. The $\Mstar=8.3\,\Msun$ track passes closest to $\Teff =
22\,500\,{\rm K}$, $\Rstar = 3.77\,\Rsun$ point, and we adopt this as the
stellar mass. With the measured rotation period (Sec.~\ref{sec:obs-reduce}) the
dimensionless angular velocity is $\omega = \Omega/\Ocrit = 0.454$,
where $\Ocrit = \sqrt{8 G \Mstar/27 \Rstar^{3}}$ is the critical angular
velocity.


\end{document}